\documentclass[12pt,preprint]{aastex}

\usepackage{textcomp}
\usepackage{color}
\usepackage{amsmath}

\def\deg      {{\ifmmode^\circ\else$^\circ$\fi}} 

\def\kmsMpc   {{km\ s$^{-1}$\ Mpc$^{-1}$}}

\newcommand{\arcsect}{\ensuremath{\mathrm{arcsec}}}
\newcommand{\arcmint}{\ensuremath{\mathrm{arcmin}}}
\newcommand{\degt}{\ensuremath{\mathrm{deg}}}

\newcommand{\Msolar}{\ensuremath{\mathrm{M}_{\odot}}}

\newcommand{\ngal}{n_\mathrm{gal}}
\newcommand{\zS}{z_\mathrm{s}}

\newcommand{\nL}{n_\mathrm{L}}

\newcommand{\rarc}{r_\mathrm{arc}}
\newcommand{\rarcsmall}{r_\mathrm{arc}^\mathrm{small}}
\newcommand{\rarclarge}{r_\mathrm{arc}^\mathrm{large}}

\newcommand{\diffd}{\ensuremath{\mathrm{d}}}

\shorttitle{Strong lenses and large-scale structure}
\shortauthors{Faure et al.}

\begin{document}


\title{On the contribution of large scale structure to strong gravitational lensing}

\author{C. Faure\altaffilmark{1,2}, J.-P. Kneib\altaffilmark{3}, S. Hilbert \altaffilmark{4,5}, R. Massey\altaffilmark{6}, G. Covone\altaffilmark{7,8}, A. Finoguenov\altaffilmark{9,10}, A. Leauthaud\altaffilmark{11}, J. E. Taylor\altaffilmark{12}, S. Pires\altaffilmark{13}, N. Scoville\altaffilmark{14}, Anton M. Koekemoer\altaffilmark{15} }

\altaffiltext{1}{Laboratoire d'Astrophysique, Ecole Polytechnique F\'ed\'erale de Lausanne (EPFL), Observatoire de Sauverny, 1290 Versoix, Switzerland}
\altaffiltext{2}{Astronomisches Rechen-Institut, Zentrum f\"{u}r Astronomie der Universit\"{a}t Heidelberg, M\"{o}nchhofstr. 12-14, 69120 Heidelberg, Germany}
\altaffiltext{3}{Laboratoire d'Astrophysique de Marseille, CNRS Universit\'e de Provence, 38 rue F. Joliot-Curie, 13388 Marseille Cedex 13, France}
\altaffiltext{4}{Argelander-Institut f\"{u}r Astronomie, Auf dem H\"{u}gel 71, 53121 Bonn, Germany}
\altaffiltext{5}{Max-Planck-Institut f\"ur Astrophysik, Karl-Schwarzschild-Strasse 1, D-85741, Garching, Germany}
\altaffiltext{6}{Institute for Astronomy, Royal Observatory, Blackford Hill, Edinburgh EH9~3HJ, U.K.}
\altaffiltext{7}{Universit\`a ``Federico II'', Department of Physical Sciences, via Cinthia 9, 80126, Napoli, Italy}
\altaffiltext{8}{Istituto Nazionale Fisica Nucleare -- Sezione di Napoli, Italy}
\altaffiltext{9}{Max-Planck-Institut f\"ur extraterrestrische Physik, Giessenbachstrasse, 85748, Garching, Germany}
\altaffiltext{10}{University of Maryland, Baltimore County, 1000 Hilltop Circle,  Baltimore, MD 21250, USA}
\altaffiltext{11}{LBNL \& BCCP, University of California, Berkeley, CA 94720}
\altaffiltext{12}{Department of Physics and Astronomy, University of Waterloo, Waterloo, Ontario N2L 3G1, Canada}
\altaffiltext{13}{Laboratoire AIM, CEA/DSM-CNRS-Universite Paris Diderot, IRFU/SEDI-SAP, Service d'Astrophysique, CEA Saclay, Orme des Merisiers, 91191 Gif-sur-Yvette, France}
\altaffiltext{14}{California Institute of Technology, MS 105-24, Pasadena, CA,91125, USA}
\altaffiltext{15}{Space Telescope Science Institute, 3700 San Martin Drive, Baltimore, MD 21218}
\begin{abstract}{
We study the correlation between the locations of galaxy-galaxy strong lensing candidates and tracers of large-scale structure from both weak lensing or X-ray emission. The COSMOS survey is a unique data set, combining deep, high resolution and contiguous imaging in which strong lenses have been discovered, plus unparalleled multiwavelength coverage. To help interpret the COSMOS data, we have also produced mock COSMOS strong and weak lensing observations, based on ray-tracing through the Millenium simulation. 
In agreement with the simulations, we find that strongly lensed images with the largest angular separations are found in the densest regions of the COSMOS field. This is explained by a prevalence among the lens population in dense environments of elliptical galaxies with high total-to-stellar mass ratios, which can deflect light through larger angles. However, we also find that the {\it overall} fraction of elliptical galaxies with strong gravitational lensing is independent of the local mass density; this observation is not true of the simulations, which predict an increasing fraction of strong lenses in dense environments. 
The discrepancy may be a real effect, but could also be explained by various limitations of our analysis. For example, our visual search of strong lens systems could be incomplete and suffer from selection bias; the luminosity function of elliptical galaxies may differ between our real and simulated data; or the simplifying assumptions and approximations used in our lensing simulations may be inadequate. Work is therefore ongoing. Automated searches for strong lens systems will be particularly important in better constraining the selection function.}
\end{abstract}


\section{Introduction}
\label{intro}

Strong galaxy lensing is a powerful tool to probe cosmological parameters. Through the angular
diameter distances involved in the lensing equation, it is possible to derive the Hubble constant
(e.g Oguri~2007; Read et al.~2007; Saha et al.~2006) and to put constraints on the
cosmological constant (e.g. Grillo et al.~2008; Chae~2003).  In practice, however,
such constraints are limited by the accuracy to which the mass distribution of the lens can be
modeled. One well-known difficulty is the mass sheet degeneracy (e.g. Gorenstein et al.~1988; Saha~2000). Moreover, the influence of the environment on the strong lensing
efficiency, and the fraction of lensing galaxies embedded in large scale structure is subject to a
lively debate (e.g. Treu et al.~2008; Auger~2008; M{\"o}ller et al. 2007; Meneghetti et al.~2007; Oguri et al.~2005; Keeton \& Zabludoff~2004; Rusin \& Tegmark~2001; Keeton et al.~1997; Kochanek~1996). Mass in the environment of a lens galaxy (nearby groups or clusters,
or intervening mass along the line of sight) is thought to (i) enhance the cross-section of strong
lensing (Fassnacht et al.~2006; Wambsganss et al.~2005; Keeton \& Zabludoff~2004), and (ii) increase the
angular separation between multiple images (Oguri et al.~2005). Therefore, failing to
take into account the environment of a modeled lens is likely to bias any cosmological results.

Until today, the main advances in understanding if and how environment influences the formation of
strong lenses have come from numerical simulations  (e.g. Wambsganss et al.~2005; Li et
al.~2005,~2006; Hilbert et al.~2007,~2008) and from statistical studies of the neighborhood of
observed lensed quasars (Moustaka et al.~2007; Williams et al.~2006; Oguri~2006; Oguri et al.~2005). Unfortunately, numerical simulations suffer from model assumptions and finite spatial
and mass resolutions; while observations are limited by the small number of strong lenses with
detailed information about their environment. No convincing answer has yet emerged on the
qualitative influence of the environment of the galaxy lens cross-section.

The COSMOS project (Scoville et al.~2007) offers a unique opportunity to compare the distribution
of strong lenses with their environment. This pan-chromatic survey covering nearly 2 square degree
on the sky has collected the richest data-set in terms of deep multi-wavelength observations,
providing unique information on the mass distribution in this field. In particular, thanks to the COSMOS HST/ACS imaging (Koekemoer et al. 2007) which has enabled the measurement
of the   weak lensing signal of distant galaxies (Leauthaud et al.~2007) and
photometric redshifts from multi-wavelength ground-based follow-up (Capak et al.~2007), Massey et
al.~(2007) have been able to reconstruct the 3D distribution of mass to redshift $z\sim1$.
Moreover, from the XMM-Newton wide field dataset (Hasinger et al.~2007), Finoguenov et al.~(2007
and 2008 in prep) have detected about 200 X-ray galaxy clusters and groups out to $z\sim1$. We
have recently (Faure et al.~2008) identified 67 strong lens candidates within the COSMOS ACS
images: one system of giant arcs in a galaxy cluster and 66 strong galaxy-galaxy lenses. The
latter include 19 galaxy-galaxy lenses with multiple images or strongly bent arcs, unambiguously
advocating their lens origin.

For the first time, we analyze the correlation between the locations of strong lenses and large
scale structure in the COSMOS field. The results are interpreted in conjunction with a simulated
survey, obtained from the Millennium Simulation (MS; Springel et al.~2005). 

The paper is organized in the following way. In $\S$~\ref{sec:sample}, we present the strong lens
candidates sample, the weak-lensing mass-map and the catalog of X-ray clusters and groups used in
this study. In $\S$~\ref{MSRT}, we introduce the Millennium Simulation, used in this paper to
compare and understand the observed distribution of strong lenses in the COSMOS field. The
correlation between the strong lens sample and the large scale environment is presented in
$\S$~\ref{lss}. The influence of the environment on the images' angular separation is studied in
$\S$~\ref{ang}. In $\S$~\ref{xray}, we investigate the correlation between the position of the
strong lenses and the X-ray groups and clusters detected in the field. The results are discussed
in $\S$~\ref{disc}. Throughout this study we assume a cosmological mean matter density
$\Omega_\mathrm{M}$ = 0.25 (in units of the critical density), a cosmological constant
$\Omega_\Lambda$ = 0.75, a Hubble constant h$_{100}$ = 0.73 (\kmsMpc), and a scale-invariant
initial density power spectrum (spectral index $n=1$) with normalization $\sigma_8=0.9$.
Magnitudes are in the AB system.

\section{Data}
\label{sec:sample}

\begin{figure*}
\begin{center}
\includegraphics[width=10.0cm]{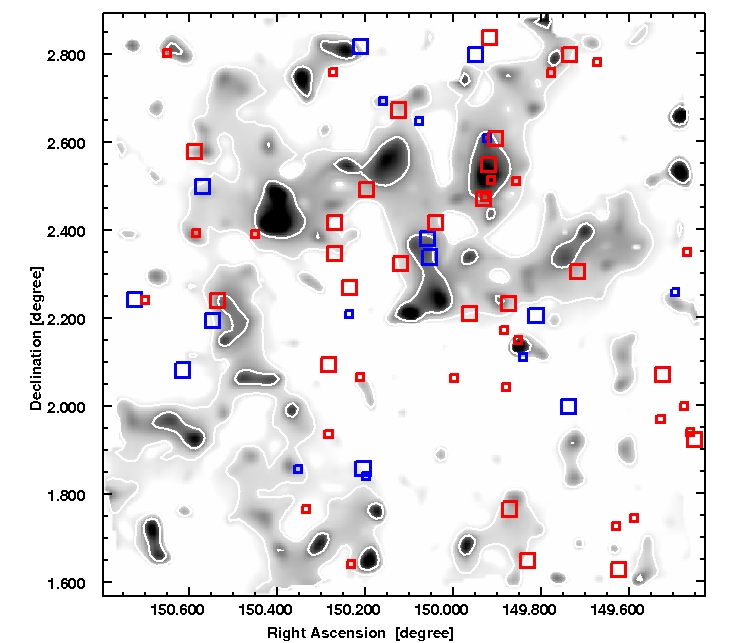}
\includegraphics[width=10cm]{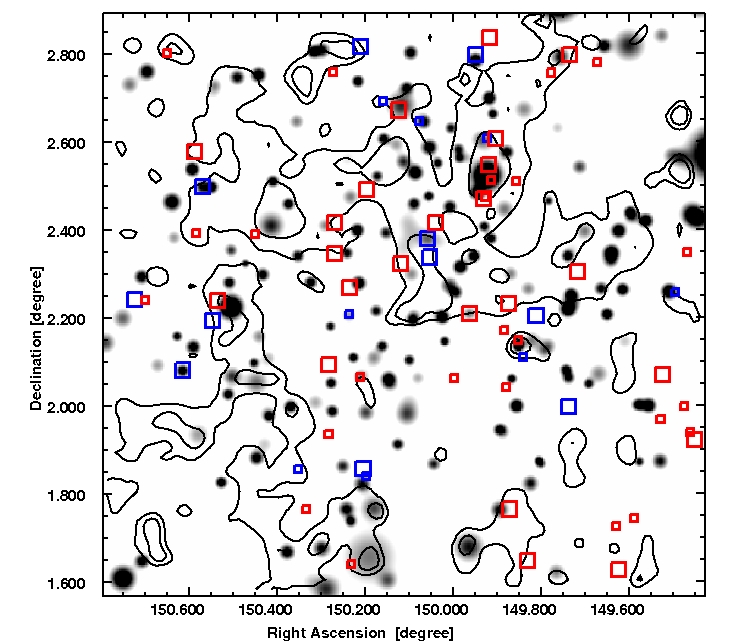}
\caption{Strong lenses projected in the COSMOS weak lensing mass map (top) and in the X-ray galaxy clusters and group mass map (bottom). Top: The linear grey-scale indicates the convergence (Massey et al.~2007), which is proportional to the (lens-efficiency weighted) projected mass along the line of sight. Bottom: the clusters/groups are represented by 2D gaussians centered at the location of the cluster/group (FWHM=2$\times$r$_{500}$, the amplitudes are scaled on their weighted surface mass-density; see $\S$~\ref{sec:sample}). Contours are drawn at $\kappa=0.4\%$ and 1.4\% in the WL mass-map and reported for comparison in the bottom map. The red squares correspond to the single arc systems and the blue squares correspond to the best systems. Small squares denote arc radii $\rarc\leq1.5\arcsec$, large squares $\rarc>1.5\arcsec$. }
\label{massmap}
\end{center}
\end{figure*}

\subsection{Strong lens sample}

In order to identify strong galaxy-galaxy lenses, Faure et al.~(2008) first selected a sample 9452 of galaxies in the COSMOS-ACS survey, which will henceforth be called the ``parent population''. This parent population consists of all galaxies in the survey with photometric redshifts $0.2\leq z_\mathrm{phot}\leq1.0$, absolute V-band magnitude $M_V<-20$, and which are classified as early-type galaxies according to their color by the photometric-redshift software (Mobasher et al.~2007).

A $10\arcsec\times10\arcsec$ region around each galaxy in the parent population was visually
inspected for multiple images and arc-like features. We also fitted and subtracted the stellar
light of the lensing galaxy, and built ground-based color images to aid the detection of genuine
strong lenses. This procedure yielded to 66 galaxy-galaxy lens candidates, which will henceforth
be referred to as simply ``lenses''. Among these 66 galaxies, 19 systems are very convincing,
showing multiple images of similar color or elongated curved arcs. They are called the ``best
systems'' throughout this paper. The other 47 systems have generally a single arc, bluer than the
lensing galaxy, and are referred to as ``single arc systems''.

\subsection{Weak lensing mass map}

The weak-lensing (WL) convergence map displayed in Fig.~\ref{massmap} (top) is a measurement of
the projected mass distribution. The convergence $\kappa$ has been calculated from the image
shapes of about half a million galaxies detected in the HST/ACS observation (Leauthaud et
al.~2007, Massey et al.~2007), which have a density $\sim70\,\arcmint^{-2}$ and a median redshift
z$\sim1.17$. The convergence map has been smoothed using a multi-scale filtering method (Starck et al.~2006) with a maximum resolution of 1.2\arcmin\, full-width at half-maximum
(FWHM). The measurement is sensitive to mass in a broad range of redshift $0.2\leq z\leq1.0$, with
a peak in the effective sensitivity at $z\approx0.7$ (Massey et al.~2007). 

Strongly lensed galaxies can typically lie at slightly higher reshifts ($z\sim 2$) due to
their additional magnification. However, the breadth of the lensing sensitivity kernel ensures
that the weak-lensing mass reconstruction spans almost the same redshift range as the strong
lens sample.

\subsection{X-ray groups and clusters}\label{sec:samplexray}

Finoguenov et al.~(2007, 2008 in prep.) have discovered 202 galaxy groups and clusters in the COSMOS field, via X-ray emission from hot gas. This techniques probes virialised structures with a mass detection threshold lower than that of the WL mass map, but with a different sensitivity as a function of redshift.
The galaxy clusters are characterized by a size $r_{500}$ (typically$\sim$0.01\deg), the radius inside which the matter density is 500 times the critical density, and where most of the cluster's X-ray flux is generated (Markevitch~1998). Fig.~\ref{massmap} (bottom) displays the X-ray galaxy clusters and groups detected in the COSMOS field. Each X-ray source is represented by
a 2D gaussian, with FWHM equal to twice the radius r$_{500}$, and amplitude scaled to the lensing efficiency of the cluster, assuming background sources at z=2. The lensing efficiency is defined by the weighted surface mass-density, $\Sigma \propto \frac{M_{500}}{\pi\times r_{500}^2}\times \frac{D_{LS}}{D_{S} }$, where  $M_{500}$ is the mass encompassed in the radius  $r_{500}$.

A comparison of Fig.~\ref{massmap} top and bottom reveals some discrepancies in the mass distribution as traced by weak lensing and X-ray emission. As noted by Massey et al.~(2007), this is mainly due to the additional sensitivity of the X-ray analysis to  compact objects. However, the weak lensing map is also noisy and potentially subject to localized defects that are not revealed by systematic tests statistically averaged over the entire field.

\section{Simulated data from the Millennium Simulation}
\label{MSRT}
\subsection{Ray-tracing setup}

To test whether our observational results are compatible with theoretical predictions, we have
also created mock COSMOS data by ray-tracing through the Millennium Simulation (MS; Springel et al.~2005). This is an $N$-body simulation of cosmic structure formation, which employed a TreePM version of \textsc{Gadget-2} with $10^{10}$ particles of mass $m_\mathrm{p}=8.6 \times 10^{8}h^{-1}\Msolar$ to follow the structure formation in a cubic region of $L=500 h^{-1}\,\mathrm{Mpc}$ comoving on a side from redshift $z=127$ and $z=0$.

The ray-tracing through the MS is based on a multiple-lens-plane algorithm described in Hilbert et
al.~(2007,~2008a,~2008b). The dark-matter distribution in the observer's backward light-cone is generated
directly from the particle data of the of the MS, and projected onto a series of lens planes. The
stellar matter in galaxies is inferred from semi-analytic galaxy-formation models implemented
within the evolving dark-matter distribution of the MS (De Lucia \& Blaizot~2007). Light rays are
traced back from an observer to their source, assuming that the rays propagate unperturbed between
lens planes, but are deflected when passing through a plane. The image distortions and
amplifications resulting from differential deflections of the light rays at the lens planes are
calculated from the projected matter distribution on the planes.

\subsection{Simulated WL mass maps}

\begin{figure*}
\begin{center}
\includegraphics[width=8cm]{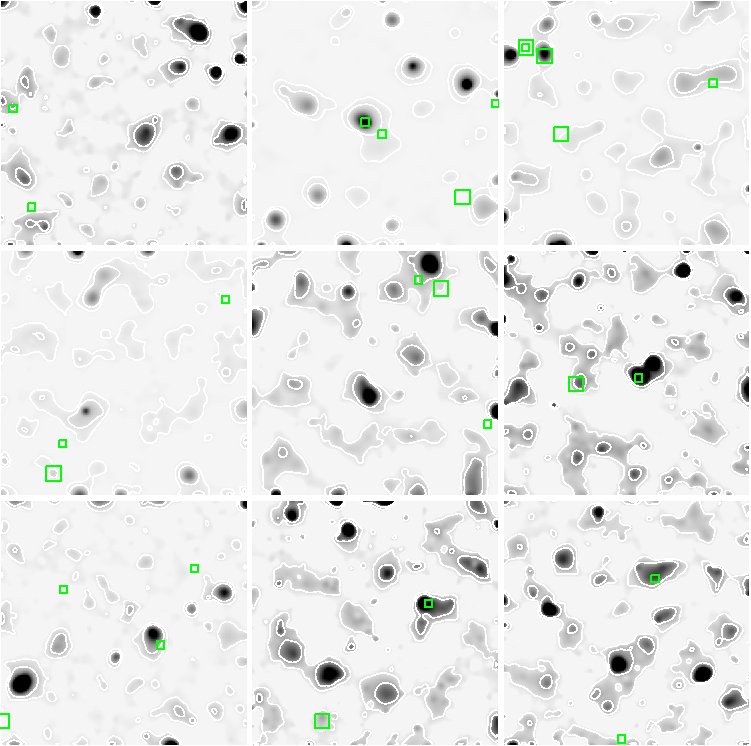}
\caption{Weak-lensing mass maps from the Millennium Simulation. Each map is 0.5\deg$\times$0.5\deg. The green squares show the location of the strong lenses. Small squares denote arc radii $\rarc\leq1.5\arcsec$, large squares $\rarc>1.5\arcsec$. Contours are drawn at $\kappa=0.4\%$ and $1.4\%$.}
\label{fig:cosmoslike}
\end{center}
\end{figure*}

We randomly choose 20 fields of $0.5\times0.5\,\degt^2$. In each field, a grid of $8192\times8192$ rays is traced through the MS by the multiple-lens-plane algorithm. We calculate the convergence $\kappa$ in each field from the ray distortions to sources at redshift $z=1.17$ (i.e.~the median redshift of the galaxies used for the weak-lensing map of the COSMOS field). We then add Gaussian noise to each pixel of the resulting convergence maps, with a variance appropriate for a source galaxy density of $70\,\arcmint^{-2}$ and galaxy ellipticity variance $\sim0.3$. The noisy convergence maps are then smoothed using the multi-scale, wavelet filtering method (Starck et al.~2006) that was applied to the real COSMOS map. This COSMOS-like WL simulation will then be processed through the same analysis as the real data, to investigate the correlation between the strong lenses and large scale structures.

In Fig.~\ref{fig:cosmoslike}, we display the projected mass map in 9 of the 20 simulated fields. 
They contain noticeably less structure than the real COSMOS data. It is already known from a study of the angular correlation of the galaxies in the COSMOS field (McCracken et al.~2007) that this field is at the upper end of the expected cosmic variance scatter. Hence we expect from the numerical simulations to give us qualitative results on the behavior of strong lenses rather than direct quantitative results that can be compared to the COSMOS field.

\subsection{Strong-lensing simulations}
From the semi-analytic galaxy catalog of De Lucia \& Blaizot~(2007), we select all galaxies with redshift $0.2\leq z \leq 1.0$, absolute V-band magnitude $M_\mathrm{V}<-20$, and bulge-to-total B-band luminosity ratio $L_\mathrm{B, bulge}/L_\mathrm{B, total}\geq0.4$ (Croton et al.~2006), in order to select similar early-type galaxies to those visual inspected in the real data. We calculate the image positions of this ``parent population'' of simulated galaxies by linear interpolation between ray positions.

The identification of strong galaxy-galaxy lenses in the COSMOS field involves the human eye, whose ``selection function'' is difficult to model. In order to model the magnification bias, we assume an apparent magnitude threshold for detecting strongly lensed galaxy images, as well as a redshift and luminosity distribution for sources in the respective filter band. From the faintest arc detected in the Faure et al.~(2008) sample, we deduce an apparent magnitude threshold of $m_\text{lim}\sim26$[mag] in the F814W filter band (i.e.~close to the COSMOS survey F814W magnitude limit; Leauthaud et al.~2007). We assume that the number density $n_\text{s}(\zS,m_\text{lim})$ of sources at redshift $\zS$ brighter than the apparent magnitude $m_\text{min}$ is simply given by a product
\begin{equation}
n_\text{s}(\zS,m_\text{lim})=p_\text{s}(\zS,m_\text{lim}) n_\text{s}(m_\text{lim})
\end{equation}
of a source redshift distribution
\begin{equation}
p_\text{s}(\zS,m_\text{lim}) = \frac{\beta \zS^2\exp\!\left[-\frac{\zS}{z_0(m_\text{lim})}\right]}{\Gamma(3/\beta) z_{0}^3(m_\text{lim})}
\end{equation}
with parameters $\beta = 3/2$ and $z_{0}(m_\text{lim}) =0.13 m_\text{lim} - 2.2$, and a cumulative source magnitude distribution
\begin{equation}
n_\text{s}(m_\text{lim})=\!\!\int\limits_{-\infty}^{m_\text{lim}}\!\!\frac{n_{0}\,\mathrm{d}m}{\sqrt{10^{2a(m_1-m)}+10^{2b(m_1-m)}}}
\end{equation}
with parameters $a=0.30$, $b=0.56$, $m_1=20$, and $n_{0} = 3\times 10^3\,\degt^{-2}$ (Smail et al.~1994, Casertano et al.~2000, Leautheaud et al.~2007, Fedeli et al.~2008). \footnote{Note that we also considered a source population with an integrated luminosity function inversely proportional to the threshold luminosity, and a source population whose images are all detected regardless of their brightness. The optical depths for such sources may differ by an order of magnitude, and the ratio between the optical depths in different weak-lensing convergence regions may be somewhat lower or higher. The qualitative behavior is, however, very similar to the case discussed here.}

Since most strong galaxy-galaxy lens candidates were selected due to arc-like features, we consider the optical depth $\tau$ for images with length-to-width ratio $r>7$ for small circular sources (see Hilbert et al.~2008a for details). Furthermore, we only consider strong-lensing regions in annuli between $0.2\arcsec$ and $5\arcsec$ around a parent-sample galaxy, since closer images could not be distinguished from the center of the lens galaxy, and more distant images are outside the visually inspected regions (Faure et al.~2008).

\section{Strong lenses as a function of weak lensing environment}
\label{lss}
In this section, we first study whether strong galaxy lenses lie preferentially in dense environments, as characterized by a high weak-lensing convergence. Then, by comparison with the simulations, we investigate which component of the large structure enhances the strong lensing cross-section. 

\subsection{Correlations in the real data}
\label{wl2}
We split the WL mass map of the COSMOS field in three regions according to their convergence values: a low-density region with $\kappa<0.4\%$, an intermediate-density region with $0.4\%\leq\kappa<1.4\%$, and the high-density region with $\kappa\geq1.4\%$ (see Fig.~\ref{massmap}). The limits between the regions are set arbitrarily, and are only aimed at studying a possible transition in the strong lenses behavior of galaxies between regions of low and high convergence. The maximum value measured in the WL map is $\kappa\sim$4.5\%. Such a low maximum convergence is due to the finite spatial resolution ($\ge$2\arcmin) of the WL mass map reconstruction, which washes out the stronger signal from the core of centrally concentrated matter structures.  For the same reason, the convergence measured in the WL mass map cannot be accounted for the convergence that directly affects the strong lenses, as what is of matter for strong lensing is the fluctuation that is coherent over the angular scale of the lens (i.e $\sim$1\arcsec). For each of the three defined convergence regions, Table~\ref{region} lists the density of elliptical galaxies and strong lenses, as well as the fraction of galaxies from the parent population that are strong lenses, best systems or single arc systems. The error bars correspond to simple Poisson statistics.

\begin{figure}[t]
\begin{center}
\includegraphics[width=8cm]{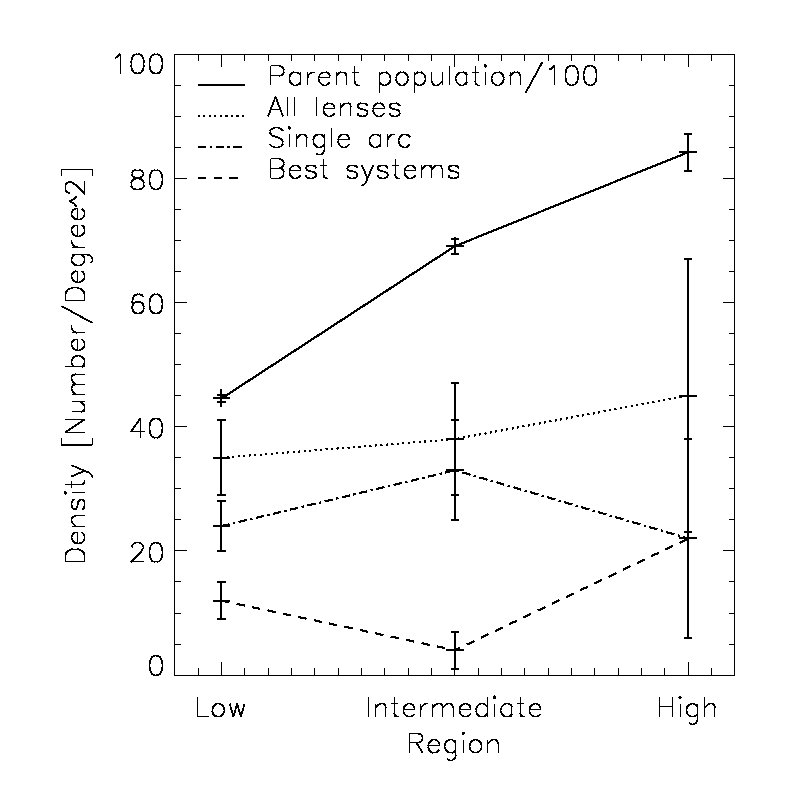}
\caption{Density of strong lens candidates as a function of the 
convergence region. The density of the parent population has been 
divided by 100 in this graph to allow comparison.\label{dense}}
\end{center}
\end{figure}

The results are summarized in Fig.~\ref{dense}. We find a comparable density of strong lenses in the low mass density environment (all lenses: $35\pm6\,\degt^{-2}$, best systems: $12\pm3\,\degt^{-2}$) and in the high mass density environment (all lenses: $45\pm22\,\degt^{-2}$, best systems: $22\pm16\,\degt^{-2}$). If we take into account the fact that the density of elliptical galaxies is about twice as high in the densest regions of the field than in the low-density region, we also find a similar fraction of elliptical galaxies that are lensing galaxies in high-density regions (all lenses: $5.4\pm2.7$\textperthousand, best systems: $2.7\pm1.9$\textperthousand) and in the low density regions (all lenses: $7.9\pm1.2$\textperthousand, best systems: $2.6\pm0.7$\textperthousand).

Surprisingly, we thus do not observe any particular correlation between the environment and the frequency of strong lenses. Strong lenses appear to be equally distributed throughout the different weak-lensing convergence regions of the field.

\begin{table*}
\renewcommand{\arraystretch}{1.1}
\centering
\begin{center}
\caption{
\label{region}
Summary of the measurements in the weak-lensing mass map.
Columns 1-3: The three convergence regions described in $\S$~\ref{wl2}.
Line 1: Covered area.
Line 2: Covered area fraction.
Lines 3-8: Number of galaxies in the parent catalog, number of all strong lens candidates, single arc systems, best systems, $\rarcsmall$ systems and $\rarclarge$ systems.
Lines 9-14: Density of galaxies, strong lenses, single arc systems, best
  systems, and $\rarcsmall$ and $\rarclarge$ systems.
Lines 15-17: Fraction of galaxies from the parent catalog which  
are strong lenses, single arc  and best systems.
}
\begin{tabular}{ l|  c c c}
\hline
Region: & Low & Intermediate & High \\
\hline
\hline
Area $[\degt^{2}]$  & 1.279  & 0.447 & 0.083   \\
\hline
Area fraction [\%] &   71      &    25     &      4.6     \\
\hline
Galaxy number& & & \\
\hspace{1em} Parent catalog                &5659 &3049 & 744 \\
\hspace{1em} All lenses                    & 45  &17   & 4   \\
\hspace{1em} Single arc systems            & 30  &15   & 2   \\
\hspace{1em} Best systems                  & 15  &2    & 2   \\
\hspace{1em} $\rarc<1.5\arcsec$ systems    &26   &6    &0    \\
\hspace{1em} $\rarc\geq1.5\arcsec$ systems &19   &11   &4    \\
\hline
Galaxy density [$\degt^{-2}$] & &  & \\
\hspace{1em} Parent catalog &4455$\pm$60     &  6806$\pm$123       &   8420$\pm $308     \\
\hspace{1em} All lenses &35$\pm$6 &38$\pm$9  &45$\pm$22\\
\hspace{1em} Single arc systems &24$\pm$4 &33$\pm$8  &22$\pm$16 \\
\hspace{1em} Best systems &12$\pm$3 &4$\pm$3 &22$\pm$16 \\
\hspace{1em} $\rarc<1.5\arcsec$  systems& $20\pm4$  & $13\pm6$ & $0\pm1$ \\
\hspace{1em} $\rarc\geq1.5\arcsec$  systems& $15\pm3$ & $25\pm7$ & $48\pm24$ \\
\hline
lens fraction [\textperthousand] & & & \\
\hspace{1em} All lenses &7.9$\pm$1.2  &5.6$\pm$1.3  & 5.4$\pm$2.7 \\
\hspace{1em} Single arc systems &5.3$\pm$1.0  &4.9$\pm$1.3  & 2.7$\pm$1.9 \\
\hspace{1em} Best systems &2.6$\pm$0.7  &0.6$\pm$0.5  & 2.7$\pm$1.9\\
\hline
\end{tabular}
\end{center}
\end{table*}

\begin{table*}
\renewcommand{\arraystretch}{1.5}
\centering
\begin{center}
\caption{
\label{sim_region}
Summary of the results from ray-tracing of 20 fields of $0.5\times0.5\,\degt^2$ through the MS. Given are the values obtained from the area of all fields combined with errors indicating the maximal deviation observed in individual fields.
Line 1: Area fraction.
Line 2: Number density $n$ of galaxies in the simulated parent catalog.
Lines 3-5: Strong-lensing optical depths $\tau$ for various source redshifts $\zS$.
Lines 6-11: As above, but restricted to systems with arc radius $\rarc<1.5\arcsec$ and $\rarc\geq1.5\arcsec$, respectively.
Lines 12-14: Strong-lensing cross-sections $\sigma$ for various source redshifts $\zS$.
Line 15: expected number $N_\text{L}$ of strong lenses (without Poisson errors) assuming a total survey area of $1.8\,\degt^2$.
Line 16: expected fraction of strong lenses among the parent population (without Poisson errors).
}
\begin{tabular}{ l| c c c}
\hline
Region: & Low & Intermediate & High  \\
\hline
\hline
Area fraction $[\%]$ & $75^{+9}_{-8} $& $19^{+7}_{-7}$ & $7^{+3}_{-2}$  \\
\hline
$\ngal$ $[\degt^{-2}]$ & $2700^{+800}_{-1200}$ & $4900^{+2400}_{-1500}$  & $8800^{+5900}_{-2500}$  \\
\hline
$\tau$ (all) $[10^{-4}]$ & & & \\
\hspace{1em}$\zS=1$  & $0.03^{+0.03}_{-0.01}$  & $0.10^{+0.16}_{-0.04}$  & $0.47^{+0.74}_{-0.32}$  \\
\hspace{1em}$\zS=2$  & $0.25^{+0.14}_{-0.13}$  & $0.74^{+0.58}_{-0.34}$  & $2.60^{+2.30}_{-1.40}$  \\
\hspace{1em}$\zS=3$  & $0.74^{+0.39}_{-0.41}$  & $2.10^{+1.80}_{-0.99}$  & $7.00^{+6.20}_{-3.70}$  \\

\hline
$\tau$ ($\rarc\leq1.5\arcsec$) $[10^{-4}]$ & & &  \\
\hspace{1em}$\zS=1$  & $0.03^{+0.02}_{-0.01}$  & $0.09^{+0.12}_{-0.04}$  & $0.26^{+0.17}_{-0.15}$  \\
\hspace{1em}$\zS=2$  & $0.20^{+0.09}_{-0.09}$  & $0.47^{+0.28}_{-0.14}$  & $1.00^{+0.78}_{-0.27}$  \\
\hspace{1em}$\zS=3$  & $0.50^{+0.20}_{-0.24}$  & $1.10^{+0.62}_{-0.32}$  & $2.30^{+1.60}_{-0.85}$  \\

\hline
$\tau$ ($\rarc>1.5\arcsec$) $[10^{-4}]$ & & &  \\
\hspace{1em}$\zS=1$  & $0.00^{+0.00}_{0.00}$  & $0.01^{+0.04}_{-0.01}$  & $0.21^{+0.57}_{-0.20}$  \\
\hspace{1em}$\zS=2$  & $0.05^{+0.05}_{-0.04}$  & $0.28^{+0.46}_{-0.22}$  & $1.60^{+1.50}_{-1.10}$  \\
\hspace{1em}$\zS=3$  & $0.24^{+0.20}_{-0.19}$  & $0.98^{+1.50}_{-0.69}$  & $4.70^{+4.60}_{-3.10}$  \\

\hline
$\sigma$ $[\arcsect^{2}]$ & & & \\
\hspace{1em}$\zS=1$  & $0.01^{+0.01}_{-0.01}$  & $0.03^{+0.03}_{-0.01}$  & $0.07^{+0.10}_{-0.05}$  \\
\hspace{1em}$\zS=2$  & $0.12^{+0.03}_{-0.02}$  & $0.20^{+0.08}_{-0.06}$  & $0.38^{+0.20}_{-0.15}$  \\
\hspace{1em}$\zS=3$  & $0.35^{+0.11}_{-0.09}$  & $0.56^{+0.20}_{-0.15}$  & $1.00^{+0.45}_{-0.43}$  \\

\hline
$N_\text{L}$ & $9^{+4}_{-4}$ & $7^{+5}_{-2}$  & $8^{+8}_{-4}$  \\
\hline
$\nL/\ngal$ [\textperthousand] & $2.4^{+1.2}_{-1.2}$ & $3.9^{+3.1}_{-1.6}$  & $7.6^{+6.4}_{-3.7}$  \\
\hline
\end{tabular}
\end{center}
\end{table*}

\subsection{Correlations in the simulated data}
\label{sec:simulations}
We now analyse our simulated data, to understand whether the absence of observed correlation between strong galaxy-galaxy lensing and weak-lensing convergence is compatible with the standard model of structure formation and galaxy evolution. The lensing simulations may also help to interpret the COSMOS observations.

The results of our lensing simulation are summarized in Table~\ref{sim_region}. We use the same threshold values for the convergence (i.e. $\kappa=$0.004 and 0.014) in the smoothed convergence maps to divide our simulated lensing fields into low-, intermediate-, and high-convergence regions as for the observed field. The resulting area fraction covered by the low-($75\pm9\%$), intermediate-($19\pm7\%$), and high-convergence regions ($7\pm3\%$) in the simulated fields is very similar to the fractions measured in the COSMOS field (except that the fraction of the simulted survey area with high weak-lensing convergence is somewhat smaller than that in the COSMOS field).

The galaxy densities $n$ of the simulated parent population in the low- and intermediate-convergence regions are slightly smaller than the densities observed in the COSMOS field. However, except for the low-convergence region, the observed densities are within the broad range spanned by the values calculated in individual ray-tracing fields. We are therefore confident that our lensing simulations reproduce the properties of the lensing observations reasonably well for a quantitative comparison of the correlation between strong galaxy-galaxy lensing and weak-lensing convergence.

\begin{figure}[t]
\centerline{\includegraphics[width=8cm]{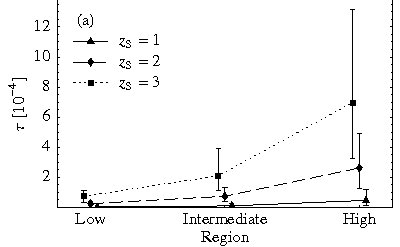}}
\centerline{\includegraphics[width=8cm]{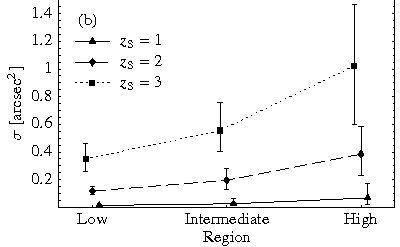}}
\caption{
\label{fig:tau_and_sigma}
Strong-lensing optical depth $\tau$ and cross-sections $\sigma$ in the low, intermediate and high weak-lensing convergence regions for sources at various redshifts $\zS$. The error bars are the dispersion of the values measured in individual fields.
}
\end{figure}

As a simple approximation, we assume that each lens galaxy induces one strongly distorted image (a ``single arc'') of a source galaxy. Thus, the density $\nL$ of strong-lens galaxies is given by an integral
\begin{equation}
\label{eq:lens_density}
\nL=\int \tau(\zS,m_\text{lim}) n_\text{s}(\zS,m_\text{lim}) \diffd\zS,
\end{equation}
of the optical depth $\tau(\zS,m_\text{lim})$ and the source density $n_\text{s}(\zS,m_\text{lim})$ for given image-detection limit magnitude $m_\text{lim}$ over all source redshifts $\zS$. In the following, we assume $m_\text{lim}=26$.

We are primarily interested in the relative number of expected strong galaxy-galaxy lenses in the different convergence regimes. Therefore, we first discuss the optical depth and take the source redshift distribution and density into account later. The optical depth $\tau(\zS)$ in the low, intermediate and high weak-lensing convergence regions is shown in Fig.~\ref{fig:tau_and_sigma}~(a) for source redshifts $\zS=1$, $\zS=2$, and $\zS=3$. There is a clear trend of an increasing strong-lensing optical depth with increasing weak-lensing convergence for all considered source redshifts. On average, the optical depth in the high-convergence region is 9-15 times larger (depending on the source redshift considered) than in the low-convergence region. This strong increase is in contrast to the weak increase of the density of strong-lens candidates seen in the COSMOS field.

The increased strong-lensing optical depth $\tau$ in regions of higher weak-lensing convergence is partly due to the increased galaxy density $n$ of the parent population. However, an increase in the strong-lensing probability remains even if the effect of the larger galaxy density is taken out. This can be seen in Fig.~\ref{fig:tau_and_sigma}~(b), where the average strong-lensing cross-section $\sigma=\tau/n$ of galaxies in the parent population is shown. The average cross-section in high-convergence regions is 3-7 times larger than in low-convergence regions. Thus, the expected fraction of strong lenses within the parent population in regions with high convergence is roughly 3-7 times larger on average than in regions of low convergence. Among individual ray-tracing fields, the ratio of cross-sections in the low- and high-convergence region ranges from 2 to 10. Hence, the simulations suggest that one should expect the fraction of strong lenses among the parent population to be {\it at least} 50 percent higher in the high-convergence regions than in the low-convergence regions (and even much larger ratios of 3-7 are expected on average).

The expected number of strong lenses in the different weak-lensing convergence regions is given by the product $N_\text{L}= A n_\text{L}$ of the area $A$ and lens density $n_\text{L}$ in the considered region. The expected fraction of galaxies in the parent population that are strong lenses is given by the ratio $\nL/\ngal$. The resulting strong-lens numbers, assuming a total survey area of $1.8\,\degt^2$, and the strong-lens fraction are shown in Table~\ref{sim_region}. 
For the high-convergence region, the predicted number and fraction of strong lenses are consistent with the values observed in the COSMOS field. However, the simulation predicts far fewer galaxy-galaxy lenses in the intermediate- and low-convergence region than there are lens candidates in the COSMOS field. \\
The reasons for the different discrepancies between observations and simulations are discussed in $\S$~\ref{versus}.

\subsection{Observations versus simulations}
\label{versus}

From  Tables~\ref{sim_region} and \ref{region}, we see that the surface covered by the three convergence regions as well as the distribution of parent population galaxies is comparable in the simulated fields and in the observations. However, the increasing fraction of strong lenses in regions of increasing convergence expected from the simulations is not matched by the constant fraction of observed strong lenses in the COSMOS field.

The discrepancy between the simulations and observations may be due to a number of reasons. Some of the possible reasons related to the observations are: 
\begin{itemize}
\item[(i)] The sample of strong lenses might be incomplete in a way that biases the results.
\item[(ii)] A significant fraction of the strong-lens candidates are not strong galaxy-galaxy lenses.
\item[(iii)] The measured WL mass map does not represent the actual mass distribution in the COSMOS field very accurately.
\item[(iv)] Our arbitrary cuts in WL convergence might not encompass important transitions between low- and high-density regimes.
\item[(v)] The COSMOS field might be a peculiar case due to cosmic variance.
\item[(vi)] The number of strong lenses might be too low to reach the necessary statistical precision, and therefore we are missing the correlation.
\item[(vii)] There may be actually no correlation between the strong lenses and the environment (at least for the used galaxy sample selection criteria).
\end{itemize}

\begin{figure}[t]
\centerline{\hspace*{-0.5cm}\includegraphics[width=9.5cm]{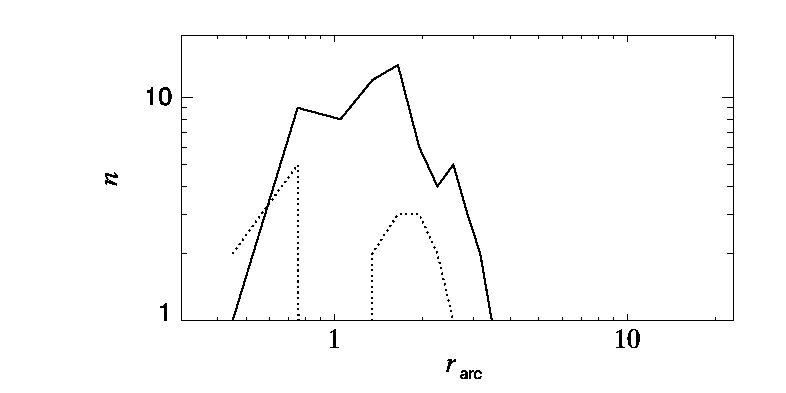}}
\centerline{\includegraphics[width=8cm]{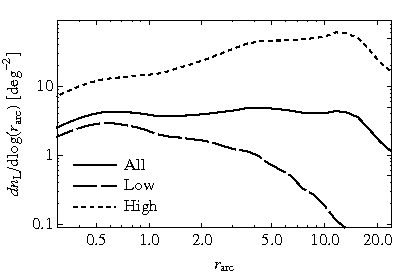}}
\caption{Distribution of arc radii in the COSMOS field (upper plot) and in the simulated fields (bottom plot). [Upper plot]: all the COSMOS lenses (solid lines), only the best systems (dotted lines). Note the lack of ``best systems'' at $r_{arc}\sim 1$ arcsec. [Bottom plot]: for the whole sky (solid lines), for the high-convergence regions (dotted lines) and for the low convergence regions (dashed lines).
\label{lograrc}}
\end{figure}

Point~(i) cannot yet be completely ruled out. Fig.~\ref{lograrc} (upper plot) shows the distribution of arc
radii in our strong lenses. Note the conspicuous absence of observed systems with arc radii around
$\sim$1\arcsec\, or greater than 3.5\arcsec. A much smoother distribution is seen in the
simulations: we would have expected to find systems with arc radii up to 5\arcsec, which is still
well within the 10\arcsec$\times$10\arcsec\, visually inspected regions. The gap at 1\arcsec\,
corresponds suspiciously to the mean seeing size of the ground based imaging. It is possible that
genuine strong lenses were excluded during the transition from monochromatic, space-based images
used to detect potential candidates, to ground-based, color images used to verify them. However, a
second independent search (Jackson~2008) did not discover any new strong lens systems amongst the
Faure et al.~(2008) elliptical galaxies, so we are confident that our sample is close to complete.
Confirming the statistical significance of the apparent lack of lenses with arc radii above
3.5\arcsec\, will require a larger sample. Overall, if the properties of strong lenses were
strongly influenced by the environment of the main lensing potential, we would have expected a
much smoother transition between systems with a galaxy-like (r$_{arc}\lesssim$4\arcsec) and
cluster-like (r$_{arc}\gtrsim$10\arcsec) mass deflectors, as seen in the simulations (Fig.~\ref{lograrc}, bottom plot). 

Point~(ii) is currently being addressed with follow-up observations (Faure et al. in preparation).
However, explaining this discrepancy would require a strong bias in the selection process as a
function of environment -- for example favoring the spurious detection of false positives in low
density environments. Point~(iii) is also subject to ongoing methodological improvements. The
insignificant reconstructed $B$-modes (Massey et al.~2007) demonstrate the overall reliability of
the weak lensing map. However, isolated imperfections in the analysis (e.g. imperfect PSF modeling
in one or two HST/ACS pointings) can go undetected when averaged over the entire field. Such
artefacts (as well as noise) can manifest as spurious convergence peaks, which would dilute the
fraction of strong lenses measured in high density environments. It is difficult for systematics
to {\it remove} real lensing convergence and thus produce the opposite effect; the much larger
area of low density than high density also means that noise acts in the same sense. This bias
doubtless contributes to the observed results, but is not at a sufficient level to explain them
alone. Point~(iv) can be excluded if the lensing simulations are not terribly unrealistic, since they
show a clear change in the strong-lens fraction. Point~(v) can also be excluded, the measured
scatter between the 20 simulated fields showing that cosmic variance alone is insufficient to
explain the discrepancy. Future observations of larger fields will also test this. They will also be required to test explanations (vi) and (vii), which cannot be ruled out with existing data.\\

Apart from possible reasons related to the observations, the discrepancy may be caused by the
simple assumptions and approximations used in the lensing simulations. These include the
assumption of circularly symmetric projected stellar-mass profiles of the lens galaxies (i.e. lensing cross-section), the neglect of the response
of their dark-matter halo to baryon cooling (Gnedin et al.~2004), the assumption of point-like
sources (instead of extended elliptical sources), and a simple source magnitude distribution
(which neglects any evolution with redshift). 
In principle, we also cannot exclude the possibility that the discrepancy is due to shortcomings of the underlying dark-matter structure formation model. Moreover, the model for the strong-lens selection function of the COSMOS sample  that we use in
the simulations might be unrealistic. Firstly, the simulated parent population relies on
semi-analytic models to predict the galaxy properties; the number of small galaxies in
high-density regions could have been overestimated. Then the detection criteria for strongly
distorted images may be too simplistic. For example, the uniform magnitude detection threshold
does not depend on the distance to the lens galaxy (apart from a cutoff for images closer than
$0.2\arcsec$), or on the lens galaxy's apparent size and brightness. In the real data, faint
lensed images at small projected distances from a large and bright lensing galaxy may have been
missed. In addition, there is considerable scatter in the average cross-sections measured in
different simulated fields (see Table~\ref{sim_region}).  A different value for the normalization $\sigma_8$  might change the total number of strong lenses, but not their distribution among the different density region.

In summary, there remain many questions (which we intend to address in future work) before we can
firmly conclude whether the observed lack of correlation between strong lenses and their
environment is real, or whether predictions from current models are right. In particular, a sample
of strong lenses identified via an automated detection algorithm would have a better constrained
selection function. This would allow further discussion of the potential inadequacies of the
simulations.

\subsection{The properties of the lens galaxies}
\label{prop}

\begin{figure}[t]
\centerline{\includegraphics[width=8cm]{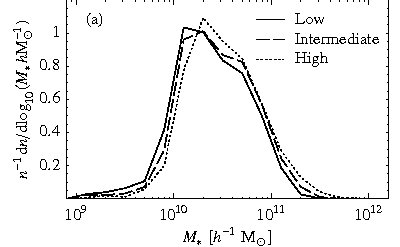}}
\centerline{\includegraphics[width=8cm]{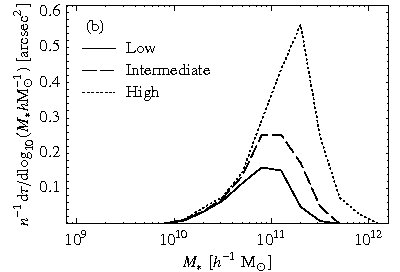}}
\caption{
\label{fig:stellar_mass_plot}
Stellar-mass distribution $n^{-1}\diffd n/\diffd \log M_*$ of the parent population and the distribution of the average lensing cross section of the parent population $n^{-1} \diffd \tau/\diffd \log M_* $ (for sources at redshift $\zS=2$) as a function of the stellar mass $M_*$ for the different weak-lensing convergence regions.
}
\end{figure}

\begin{figure}[t]
\centerline{\includegraphics[width=8cm]{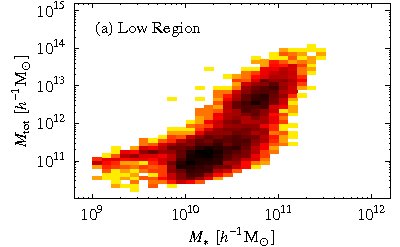}}
\centerline{\includegraphics[width=8cm]{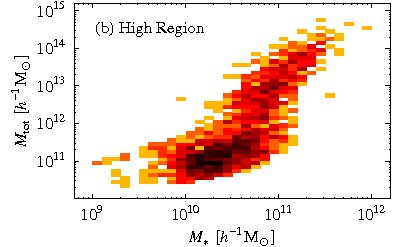}}
\caption{
\label{fig:stellar_and_total_mass_plot}
Distribution of the stellar mass $M_*$ and total mass $M_\text{tot}$ of the parent population for the low-convergence regions (a) and the high-convergence regions (b).
}
\end{figure}

\begin{figure}[t]
\begin{center}
\includegraphics[width=8cm]{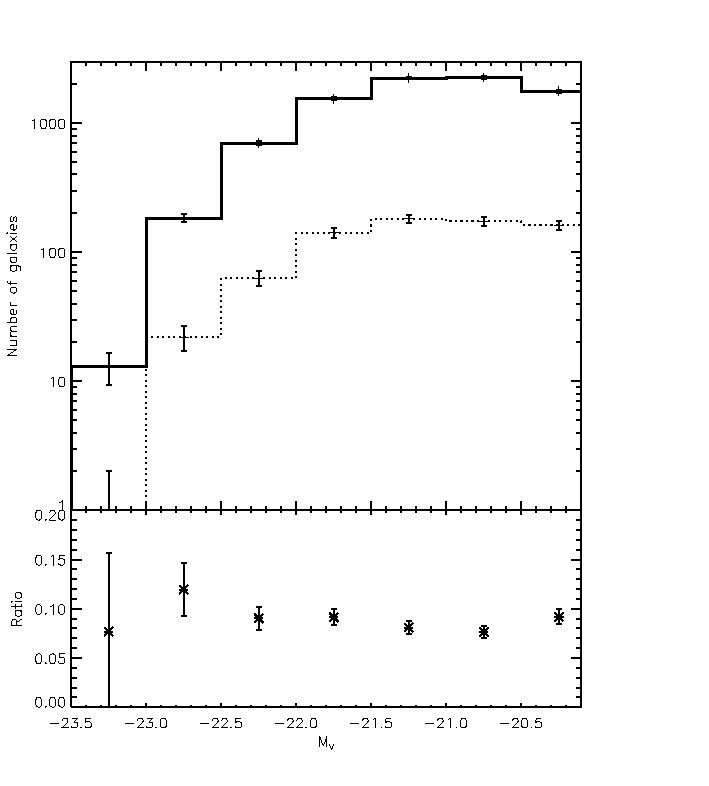}
\caption{
\label{histomv}
Distribution of elliptical galaxies from the parent catalog versus their absolute magnitude in the V-band. Top panel: Solid line histogram: galaxies in the low+intermediate convergence regions. Dashed line: galaxies in the high convergence level region. Error bars are poisonnian. Bottom panel: Ratio between the histograms values.
}
\end{center}
\end{figure}

\begin{figure}[t]
\centerline{\includegraphics[width=8cm]{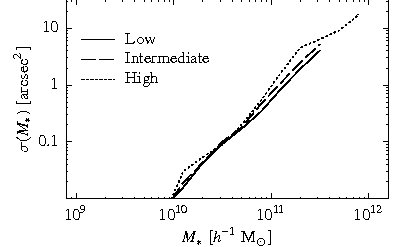}}
\caption{
\label{fig:stellar_mass_sigma_plot}
Cross-section $\sigma(M_*)$ of galaxies with stellar mass $M_*$ for strong lensing of sources at redshift $\zS=2$ for the different weak-lensing convergence regions.
}
\end{figure}

The semi-analytic galaxy catalogue (De Lucia \& Blaizot~2007) used for the lensing simulations allow us to investigate the relation of various galaxy properties to observed trends in the strong lensing probabilities. Fig.~\ref{fig:stellar_mass_plot}(a) shows the distribution $n^{-1} \diffd n/\diffd \log M_*$ of the stellar mass $M_*$ in galaxies of the parent population, for the three convergence regions. The distributions are very similar apart from a small shift towards higher stellar masses for regions with larger weak-lensing convergence. This shift, however, implies a higher number of galaxies with stellar masses $M_*>10^{11}h^{-1}\Msolar$. Although a small fraction of the entire parent populations ($\sim$4\% of the parent population in low-convergence regions, $\sim$9\% in high-convergence regions) these galaxies contribute significantly to the average lensing cross-section of galaxies, as seen in Fig.~\ref{fig:stellar_mass_plot}(b). These massive galaxies not only have higher stellar masses (and thus higher luminosities), but are often associated with significantly more massive dark-matter halos associated with them, resulting in very large total-to-stellar-mass ratios $M_\text{total}/M_\text{stellar}>100$ (and large mass-to-light ratios), as can be seen in Fig.~\ref{fig:stellar_and_total_mass_plot}. There are only a few of them, so they do not significantly affect the luminosity distribution of the parent population, but they dominate the lensing statistics.

This result of the simulation is not in contradiction with the observations in the COSMOS field: the brightest galaxies do not dominate the population of galaxies in the high-convergence region (Fig.~\ref{histomv}). To verify the existence of these high total-to-stellar mass galaxies, galaxy-galaxy- weak lensing analysis of the COSMOS galaxies will be necessary (such as Mandelbaum et al.~2005), as well as a detail mass model of the strong lenses of the sample.
Nevertheless, the low correlation between the strong lenses and the mass map measured in the COSMOS field suggests that there are fewer of theses massive galaxies in the observed field rather than in the simulated fields. This result, if verified, could also pinpoint a problem in the semi-analytic model used to represent the galaxy luminosity function in the simulations.

The higher cross-section in the high-convergence regions predicted by the simulations is not only due to the higher number of massive galaxies. In Fig.~\ref{fig:stellar_mass_sigma_plot}, the (average) strong-lensing cross-section $\sigma(M_*) = (\diffd \tau/ \diffd M_*)/(\diffd n/\diffd M_*)$ as a function of stellar mass $M_*$ is shown for the different weak-lensing convergence regions. Even if we compare galaxies at fixed stellar mass $M_*$, the cross-section $\sigma(M_*)$ is consistently larger in the high-convergence regions than in the low-convergence regions. Using the simulation data, we can identify at least one cause: For given stellar mass, galaxies in high-convergence regions tend to have larger dark-matter halos than galaxies in low-convergence regions. However, other possible causes, e.g. additional matter from group/cluster halos or other structures along the line-of-sight, cannot be excluded so far (and should be investigated in more detail in future work).

\section{Arc radii as a function of weak lensing environment}
\label{ang}

The images of a source deflected by a gravitational lens are formed at stationary points of the arrival-time surface. The image positions are therefore directly related to the convergence and the associated shear of the total lens potential. One may conjecture that lens galaxies in high-convergence regions will generate larger image separations for multi-image systems than lens galaxies in low-convergence regions. Using a model derived from $N$-body simulations and galaxy formation models, Oguri et al.~(2005) studied the impact of external convergence and shear on the image-separation distribution of quasars that are multiply imaged by galaxies. They found that the angular separation between multiple images is strongly correlated with the surface mass density of the environment and that the external convergence enhances the lensing probability by $\sim30\%$ for systems with arc radius of $\rarc\approx1.5\arcsec$ (and up to 200\% for $\rarc\approx2.5\arcsec$). 

\subsection{Arc radii in the COSMOS field}

\begin{figure}[t]
\begin{center}
\includegraphics[width=8cm]{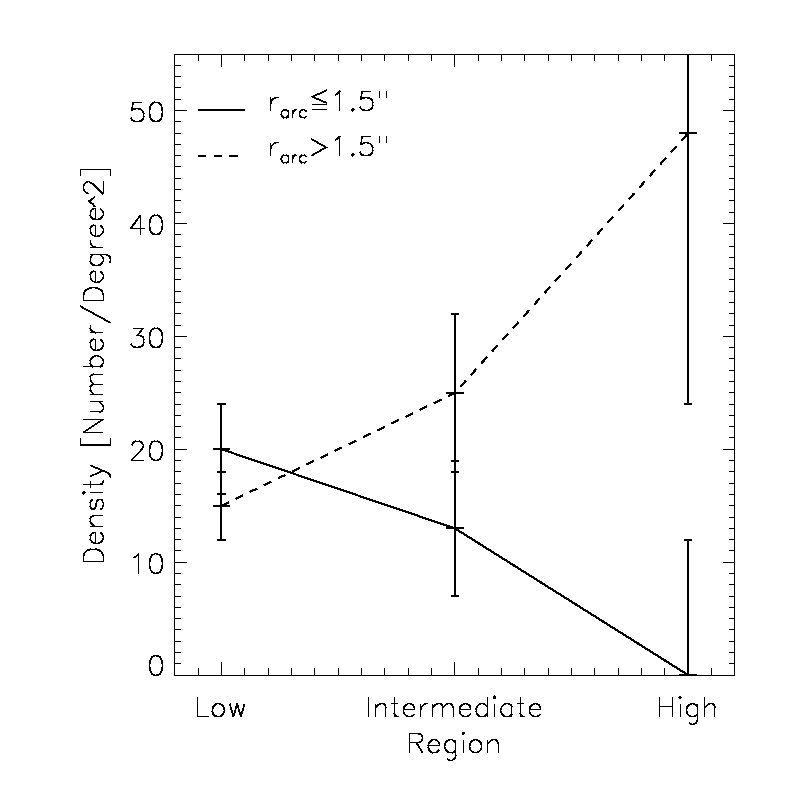}
\caption{Density of strong lens candidates (in $\degt^{-2}$) in the 
three convergence regions according to the arc radius. \label{arc}}
\end{center}
\end{figure}

Here we want to verify whether the distribution of strong galaxy-galaxy lenses arc radii in the
COSMOS field is a function of the environmental convergence. For that purpose, we split the lens
sample into two (similarly sized) sub-samples according to their arc radii: $\rarcsmall$ if
$\rarc<$1.5\arcsec\, and $\rarclarge$ if $\rarc\geq1.5\arcsec$. The largest angular separation in
our sample is $\rarc=3.54$\arcsec. The number and density of strong lenses as a function of
environment are listed for the two sub-samples in Table~\ref{region}, and their densities are
plotted in Fig.~\ref{arc}. We find that the densities of $\rarcsmall$ and $\rarclarge$ systems are
similar in low- and intermediate-convergence regions. However, only $\rarclarge$ systems are
found in high-convergence environments ($48\pm24\,\degt^{-2}$) and their density is up to
four times higher than in low-convergence regions ($15\pm3\,\degt^{-2}$). 

\subsection{Arc radii in the simulated fields}

\begin{figure}[t]
\centerline{\includegraphics[width=8cm]{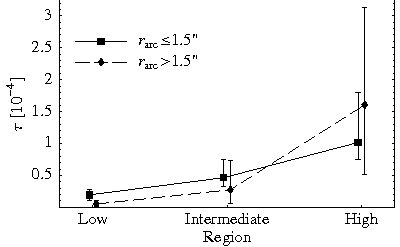}}
\caption{
\label{fig:tau_small_large_r}
Strong-lensing optical depth $\tau$ for sources at redshift $\zS=2$ in the low, intermediate and high weak-lensing convergence regions for systems with arc radius $\rarc<1.5\arcsec$ and for systems with $\rarc\geq1.5\arcsec$. The error bars indicate the observed range of values in individual fields.
}
\end{figure}

To estimate the density of strong-lens systems with small and large arc radii, we separately
calculate the strong-lensing optical depth for systems with arc radii $\rarc<1.5\,\arcsec$ and for
systems with $\rarc\geq1.5\,\arcsec$. The resulting optical depths are shown in
Fig.~\ref{fig:tau_small_large_r} and given in Tab \ref{sim_region} for different source redshifts.
The optical depth for $\rarclarge$ systems increases very strongly with increasing weak-lensing
convergence ($\sim$20 times higher in high-convergence regions). This matches the observed data,
and is due to a combination of external shear/convergence plus an increased fraction of high
mass-to-light ratio galaxies in dense environments. 

The optical depth for $\rarcsmall$ systems also increases ($\sim$ 5 times higher in
high-convergence regions) in the simulations. An equivalent increase in the number of observed
$\rarcsmall$ systems is not seen in the COSMOS data. However, as already mentioned in
$\S$~\ref{versus}, faint lensed images at small projected distances from a large and bright lens
galaxy may not be visible in real images, thereby leading to an underestimate of the number of
small arc radius systems.

\subsection{Conclusions on the arc radius distribution}

High-density environments enhance the production of strong lens systems with large image
separations with respect to low-density environments. This effect is observed both in the
simulations and in the COSMOS field. The large image separations are due to the presence of
high mass-to-light ratio galaxies in the dense regions and the strong external shear/convergence
generated by the lens-galaxy environment. The latter can also be induced by massive structures along the same line-of-sight.

As discussed in $\S$~\ref{prop}, we do not find substantial evidence from the galaxy luminosity
function (Fig.~\ref{histomv}) that the brightest galaxies (also likely to be the most massive)
are preferentially found in high density environments. Nor does the low (constant) fraction
of strong lenses in high-convergence regions of the COSMOS field hint at a excess of massive galaxies. However, the presence of only strong lenses with large image separations implies that this is indeed the case; or simply that it is only external shear that increases the image separation, not the presence of massive galaxies.

\section{Strong lenses as a function of X-ray environment}
\label{xray}

As is apparent in Fig.~\ref{massmap}, the WL mass map and the $X-$ray galaxy clusters and groups do
not trace exactly the same mass distribution, the latter showing peaks of mass with low spatial
extension missed by the WL mass map. In order to probe the correlation of strong lenses with their
environment at higher resolution than the weak lensing mass map, we have studied the correlation
between strong lenses and their environment as traced by $X-$ray emission (see
$\S$~\ref{sec:samplexray}).

\subsection{Lensing galaxies lying in the cluster/group center}

Seven lensing galaxies (2 ``best system'' and 5 ``single arc'' systems) are members of an $X-$ray
galaxy cluster or group (lying inside a projected radius $r_{500}$ and within redshift $\Delta
z=\pm0.05$). These represent 10$\pm$4\% of the strong lenses (10$\pm$7\% of the ``best systems'').

Two lensing galaxies (1 ``best system'' and 1 ``single arc''; both $\rarclarge$ systems) are
within only $0.02~r_{500}$ of the cluster core. The other five lensing galaxies (of which two are
$\rarclarge$ systems) lie at projected distances $>0.3~r_{500}$ from the cluster center.

\subsection{Lensing galaxies aligned with a galaxy cluster}

We find that 14 systems (6 ``best systems'' and 8 ``single arc'' systems) have an impact parameter with a galaxy cluster smaller than $r_{500}$, but have a different redshift than the cluster. By adding the 7 systems embedded in a galaxy cluster, we find that 31$\pm$7\% of strong lenses (and 42$\pm$15\% of the best systems) are within or aligned with a galaxy cluster, with an impact parameter inside $r_{500}$. In comparison, Williams et al.~(2006), in a study of the environment of 12 strongly lensed quasars with similar arc radii to our sample, found that 50$\pm$20\% to 67$\pm$23\% of strong lenses are aligned with a galaxy cluster/group. Within uncertainties, we are thus in rough agreement with their results.

\begin{figure}[t]
\begin{center}
\includegraphics[width=8cm]{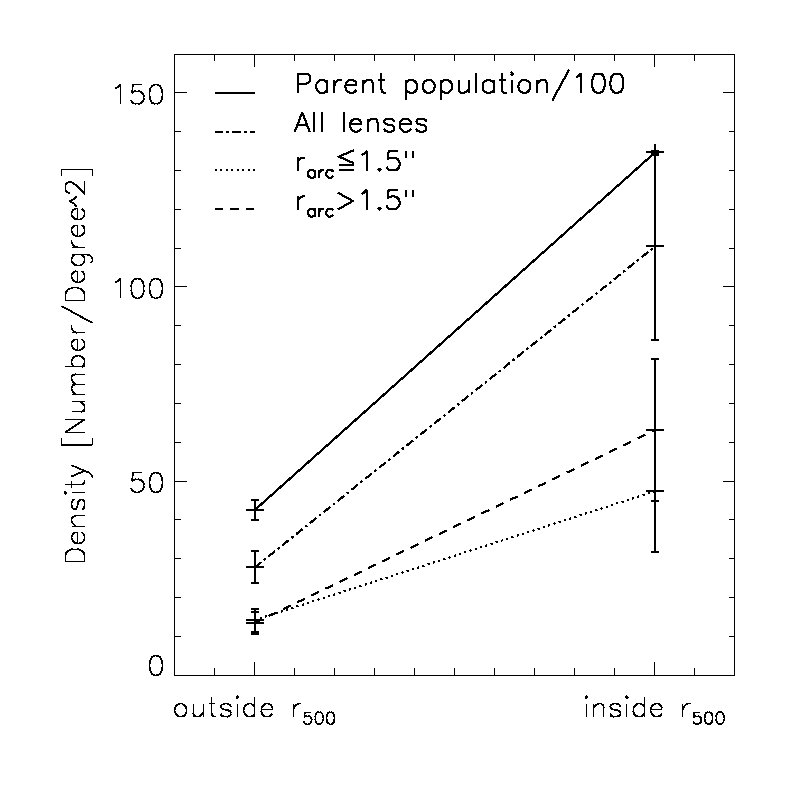}
\caption{Density of strong lenses off-center with the central region of a cluster (outside $r_{500}$ ) and aligned with the central region of a galaxy cluster/group, or in it at a distance $r_{500}\leq1$. The density of the parent population has been divided by 100 in this graph to allow comparison. 
\label{xproj}}
\end{center}
\end{figure}

If we assume that galaxy clusters in the COSMOS field are circular, do not overlap and have a
projected size of $\pi  r_{500}^2$, they cover a projected area of $0.190\,\degt^2$. We find a
density of $110\pm24\,\degt^{-2}$ strong lenses inside this area, and a density of
$28\pm4\,\degt^{-2}$ outside it (see Fig.~\ref{xproj}). The density of strong lenses is thus
$\sim$4 times higher in the projected region covered by galaxy clusters. However, the density of
galaxies from the parent population (Fig.~\ref{xproj}) is also $\sim4$ times higher in this
region. The fraction of elliptical galaxies that are strong lenses is therefore approximately
constant, regardless of whether the lensing galaxy is enclosed within (6$\pm$1\textperthousand) a
projected radius $r_{500}$ from the center of a galaxy cluster or groups or not
(7$\pm$1\textperthousand). This recovers the result of $\S$~\ref{lss}, that the population of
strong lenses follows the distribution of bright elliptical galaxies, and that additional matter
structures along the line-of-sight do not significantly increase the strong lensing cross-section.

The density of $\rarcsmall$ and $\rarclarge$ systems is similar in the regions away from lines of
sight to clusters ($14\pm3\,\degt^{-2}$ and $11\pm3\,\degt^{-2}$ respectively). However, the
density of $\rarclarge$ systems is slightly higher ($63\pm18\,\degt^{-2}$ for $\rarclarge$ and
$47\pm16\,\degt^{-2}$ for $\rarcsmall$ systems) when the projected distance of a lensing galaxy to
a galaxy cluster center is within $r_{500}$. As in $\S$~\ref{ang}, two explanations are possible:
additional matter along the line of sight increases the angular separation between the images,
and/or more massive galaxies are found in clusters, which boost the lensing cross-section. 

\begin{figure}[t]
\begin{center}
\includegraphics[width=8cm]{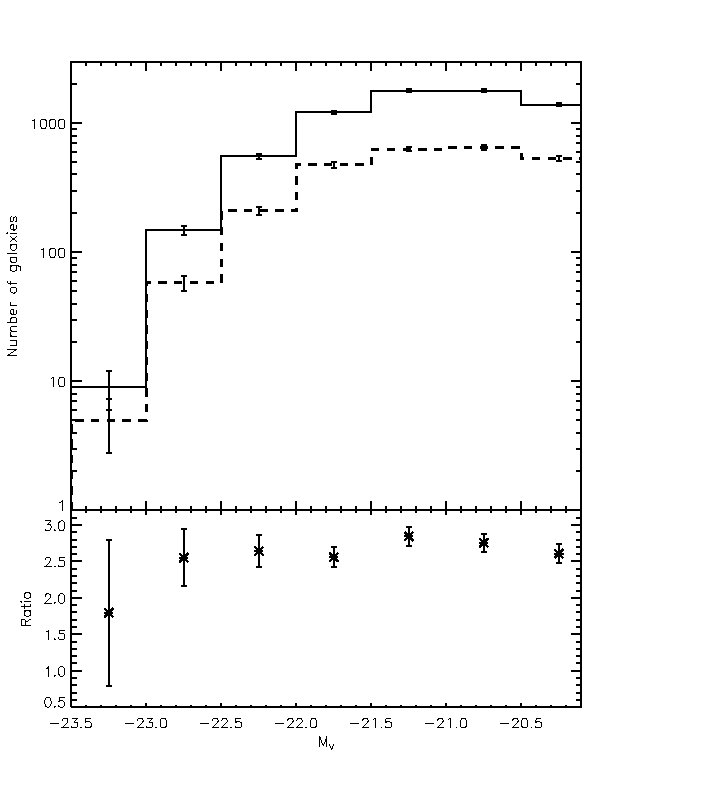}
\caption{ Distribution of elliptical galaxies from the parent population versus their absolute-magnitude in the V-band. Top panel: Dashed line histogram: the distribution of galaxies in the region covered by the galaxy clusters (in $r_{500}$). Solid line histograms: the distribution of galaxies outside this region. Bottom panel: ratio of these two distributions
\label{mvelli}}
\end{center}
\end{figure}

Fig.~\ref{mvelli} shows the distribution of elliptical galaxies in the parent population, according to their brightness and as a function of their location. We find a similar distribution of galaxies inside and outside the region covered by clusters. An excess of bright, massive galaxies is therefore not a plausible explanation for the excessive fraction of $\rarclarge$ in the region covered by the clusters, unless the galaxies in this region have an intrinsically higher mass-to-light ratio than those outside. 

\section{Summary and discussion}
\label{disc}

We reach the same conclusion by studying the correlation between strong gravitational lenses and
large scale structure, as traced by both weak-lensing convergence and $X-$ray emission. We find
that dense environments do not enhance the fraction of galaxies that act as strong lenses (with
image separations r$_{arc}\lesssim$4\arcsec), but do increase the number of lenses with large
image separations.  

We have constructed mock observations of gravitational lensing by ray-tracing through the
Millennium Simulation (Springel et al.~2007). Simulated regions of high weak-lensing convergence
similarly increase the number of strong lenses with large image separations. However, in contrast
to our observations, the simulations also predict a higher {\it total} fraction of strongly
lensing galaxies in high convergence regions than in low convergence regions. Several possible
reasons for this discrepancy are discussed in $\S$~\ref{versus}, including imperfect modeling of
the parent galaxy sample, or possible bias in the strong lens selection function. For example, the
apparent magnitude and redshift distribution of the source galaxy population, in particular for
high redshifts and faint magnitudes, should be improved, since these are crucial for modeling the
magnification bias in strong lensing observations. Nor did we  take into account the shapes and
sizes of sources, images or lens galaxies, and may have thus missed real strongly lensed images
close the lens center (Meneghetti et al.~2008). In future surveys, automatic detection software,
such as $Arcfinder$ (Seidel \& Bartlemann~2007) or $Haggles$ (Marshall et al.~2008), will be
important to quantify the selection bias. This should ease the comparison between observations and
simulations. We will investigate such possibilities in a further paper (Faure et al.\, in
preparation).

According to the Millennium Simulations, the main cause for the higher fraction of strong lens
systems with large image separations in high-density regions is an excess of galaxies with high
total-to-stellar mass ratio. However, neither the existence of such galaxies, nor the fraction of
galaxies that they constitute within the field is entirely clear. Detailed mass modeling of the
various galaxy populations via galaxy-galaxy weak lensing analysis (Leauthaud et al. 2008 in
preparation) will help explore the existence, location and lensing properties of such massive
galaxies.

Constraining simulations and cosmological parameters will ultimately require a larger sample of
strong lenses. However, to understand the effect of local environment, it is vital that the
environment be well measured: a task for which the COSMOS survey has been uniquely well designed.
The CFHTLS Strong Lensing Legacy Survey (SL2S, Cabanac et al.~2007) also aims to explore the
relation of strong-lenses to their environment, but is only sensitive to larger Einstein radius
systems. The final word may only come from the proposed SNAP/JDEM mission, which will probe a full
range of arc radii.

%

\acknowledgments
We are  thankful to the referee for its useful report.
 JPK acknowledges support from CNRS, CNES and the ANR
through the grant 06-BLAN-0067. SH is supported by the DFG within the
Priority Programme 1177 under projects SCHN 342/6 and WH6/3. RM is
supported by STFC Advanced Fellowship PP/E006450/1.



\end{document}